\documentclass{article}

\usepackage{microtype}
\usepackage{graphicx}
\usepackage{subcaption}
\usepackage{booktabs} 

\usepackage{hyperref}

\usepackage[accepted]{icml2026}

\usepackage{amsmath}
\usepackage{amssymb}
\usepackage{mathtools}
\usepackage{amsthm}
\usepackage{comment}
\usepackage{multirow}
\usepackage{booktabs}
\usepackage{makecell}
\usepackage[table]{xcolor} 
\usepackage{xcolor}
\usepackage{makecell}
\usepackage{amssymb}
\usepackage{pifont}
\usepackage{amsmath}

\usepackage[capitalize,noabbrev]{cleveref}

\theoremstyle{plain}

\theoremstyle{definition}

\theoremstyle{remark}

\usepackage[textsize=tiny]{todonotes}

\icmltitlerunning{Focus Then Listen}

\begin{document}

\twocolumn[
  \icmltitle{Focus Then Listen: An Empirical Study of Plug-and-Play Audio Enhancer for Noise-Robust Large Audio Language Models}



  \icmlsetsymbol{equal}{*}

  \begin{icmlauthorlist}
    \icmlauthor{Han Yin}{kaist}
    \icmlauthor{Yang Xiao}{melb}
    \icmlauthor{Younghoo Kwon}{kaist}
    \icmlauthor{Ting Dang}{melb}
    \icmlauthor{Jung-Woo Choi}{kaist}
  \end{icmlauthorlist}

  \icmlaffiliation{kaist}{School of Electrical Engineering, KAIST, Daejeon, Republic of Korea}
  \icmlaffiliation{melb}{University of Melbourne, Australia}

  \icmlcorrespondingauthor{Jung-Woo Choi}{jwoo@kaist.ac.kr}

  \icmlkeywords{Machine Learning, ICML}

  \vskip 0.3in
]



\printAffiliationsAndNotice{}  

\begin{abstract}
  Large audio language models (LALMs) are a class of foundation models for audio understanding. Existing LALMs tend to degrade significantly in real-world noisy acoustic conditions where speech and non-speech sounds interfere. While noise-aware fine-tuning can improve robustness, it requires task-specific noisy data and expensive retraining, limiting scalability. To address this issue, we propose Focus-Then-Listen (FTL), a plug-and-play audio enhancer that improves LALMs' noise robustness. Specifically, FTL first separates the input waveform into speech and non-speech, and a modality router is applied to predict the target audio modality (e.g., speech) based on the user's instruction. Finally, a modality-aware fusion block generates a task-adaptive enhanced signal for improved downstream perception and reasoning. Experiments across multiple LALMs and tasks show that FTL improves performance across different noise levels without fine-tuning on LALMs.
\end{abstract}

\section{Introduction}
Large audio language models (LALMs) have recently emerged as a powerful paradigm for unified audio understanding and reasoning~\cite{qwen-audio,gama,audio-reason}. By integrating audio perception with large language models (LLMs), LALMs enable a wide range of applications, including speech recognition, acoustic scene analysis, and audio question answering~\cite{lalm-asr, pengi, audiobench}. 

Noise robustness remains a fundamental challenge for LALMs~\cite{li2026silence}. Here, the noise refers to acoustic signals that are irrelevant to the user's intent in a given task. For instance, in speech understanding tasks, non-speech sounds can be the noise, whereas in environmental sound analysis, speech may act as interference. In real-world environments, audio inputs are rarely clean and often contain multiple overlapping or irrelevant components. Without sufficient robustness to such task-irrelevant signals, LALMs may misinterpret the user's intent, resulting in degraded interaction quality and unreliable system behavior, particularly in safety-critical applications~\cite{voice_smart_home,in-car-voice,robot-public}.

\begin{figure}[t]
\centering
\includegraphics[width=1\columnwidth]{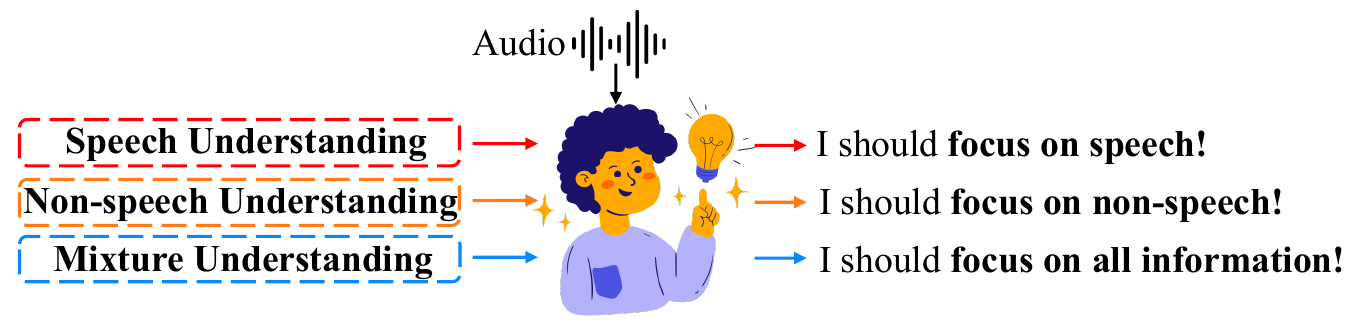}
\caption{Process of human audio understanding.}
\label{fig1:intro}
\vspace{-2em}
\end{figure}

\begin{figure*}[t]
\centering
\includegraphics[width=1.5\columnwidth]{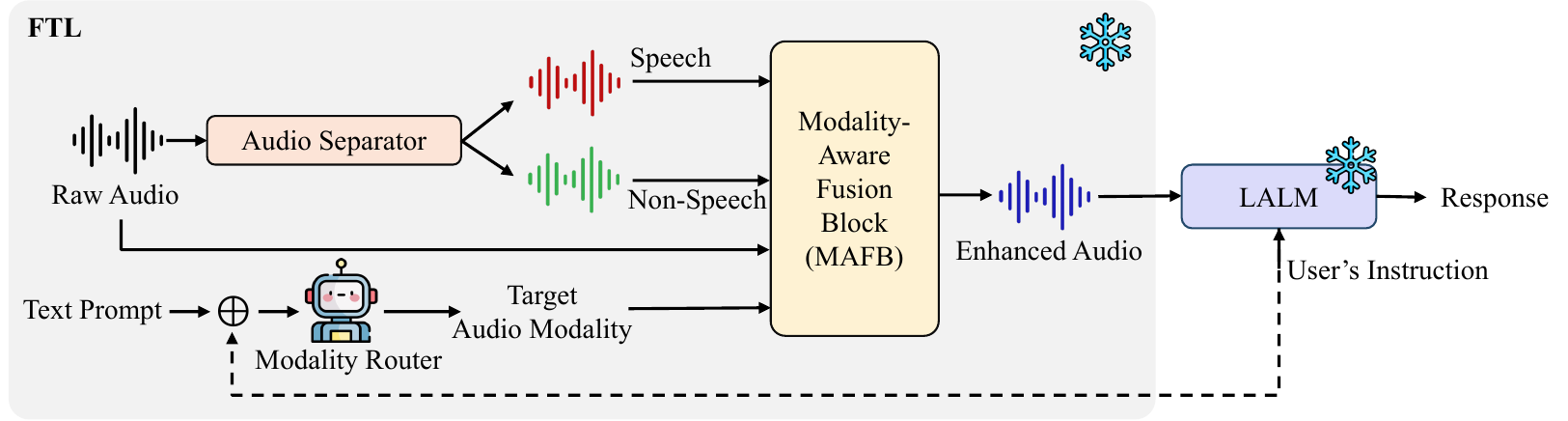}
\caption{Overview of proposed audio enhancer (FTL) for noise-robust
large audio language models.}
\label{fig2:overview}
\vspace{-1.5em}
\end{figure*}

Recent work has begun to investigate this problem. SSEU-Bench~\cite{sseu-bench} explicitly models the coexistence of speech and non-speech sounds and considers their energy imbalance across diverse scenarios. An important observation is that cross-component interference significantly affects model performance: when performing speech understanding, strong non-speech sounds can degrade recognition, and similarly, dominant speech can negatively impact non-speech sound understanding. To address this issue, SSEU-Bench uses chain-of-thought (CoT) prompting to decompose complex audio understanding into simpler steps. However, the improvement is mainly observed in audio tagging tasks, and CoT often requires task-specific prompt design. Another straightforward approach to enhance robustness is noise-aware training, which involves fine-tuning models on large-scale datasets augmented with various noise types~\cite{noise-lalm-asr, kimi}. This paradigm requires extensive data curation, as covering the infinite variability of real-world noise is practically infeasible. In addition, fine-tuning may also lead to catastrophic forgetting or degrade performance on clean data~\cite{empirical,zhai2023investigating,yin2015noisy}. In SEE~\cite{zhang2026see}, researchers propose an embedding-based approach for developing noise-robust LALMs, but assumes that noise is explicitly pre-defined (e.g., Gaussian noise) and the isolated pure-noise recordings are required during training. This assumption is incompatible with our setting, where noise cannot be pre-defined but is task-dependent: non-speech acts as noise for speech tasks, and vice versa.

To address these issues, we explore to use an audio enhancer, i.e., Focus-Then-Listen (FTL), to improve LALMs' noise robustness. 
Our motivation stems from the human audio understanding process.
As illustrated in Fig.~\ref{fig1:intro}, when confronted with audio, humans selectively focus on the component relevant to their intent. 
Inspired by this, FTL infers the task-relevant audio modality from the user’s instruction and produces a filtered, modality-aligned signal for the LALM, which improves downstream perception and reasoning in noisy conditions. Our contributions are as follows:
\begin{itemize}
    \item We propose FTL, a plug-and-play audio enhancer that improves LALMs' noise robustness. To the best of our knowledge, this is the first work to explore mitigating speech and non-speech interference for LALMs via instruction-aware audio enhancement. 
    \item Experiments across multiple LALMs and benchmarks demonstrate the effectiveness of the proposed FTL in both audio perception and reasoning tasks\footnote{\url{https://sites.google.com/view/ftl-lalm}}\label{fn:page}.
\end{itemize}

\section{Focus Then Listen}
\subsection{Overview} 

As shown in Fig.~\ref{fig2:overview}, in FTL, we first use an audio separator to separate the raw input audio into speech and non-speech tracks, which can be expressed as:
\begin{equation}
    \mathbf{S_{sp}}, \mathbf{S_{ns}} = Sep(\mathbf{S_{ra}})
\end{equation}
where $Sep(\cdot)$ is the audio separator, $\mathbf{S_{ra}}$ is the raw audio, $\mathbf{S_{sp}}$ and $\mathbf{S_{ns}}$ are the separated speech and non-speech, respectively.
We then introduce a modality router to determine the target audio modality $m$ based on the user's instruction. If the task only requires speech-related information, the router should output ``\textit{speech}''; if the task focuses on non-speech content, the router outputs ``\textit{non-speech}''. For more complex tasks that require both modalities, the router outputs ``\textit{mixture}''. Specifically, we use an LLM as the modality router; the detailed prompt used for the LLM is provided on Appendix~\ref{prompt:agent}. 
Finally, we employ a modality-aware fusion block (MAFB) to generate task-adaptive enhanced audio conditioned on the selected modality. The goal of the MAFB is to refine the acoustic signal to better align with the user's task.
Through FTL, we aim to amplify task-relevant information while suppressing irrelevant components in the audio, allowing the downstream LALM to focus more effectively on informative acoustic cues.

\subsection{Modality-Aware Fusion Block}

The MAFB is designed to generate task-adaptive enhanced audio based on the modality selected by the router. Specifically, the enhanced audio $\mathbf{S_{en}}$ is computed as:
\begin{equation}
\resizebox{0.9\columnwidth}{!}{$
\mathbf{S_{en}} =
\begin{cases}
\alpha_{sp} \mathbf{S_{sp}} + (1-\alpha_{sp})\mathbf{S_{ra}}, & \text{if } m = \text{``\textit{speech}''} \\
\alpha_{ns} \mathbf{S_{ns}} + (1-\alpha_{ns})\mathbf{S_{ra}}, & \text{if } m = \text{``\textit{non-speech}''} \\
\mathbf{S_{ra}}, & \text{if } m = \text{``\textit{mixture}''}
\end{cases}
$}
\label{eq:fusion}
\end{equation}
where $m$ denotes the target audio modality predicted by the router. The coefficients $\alpha_{sp}$ and $\alpha_{ns}$ are hyperparameters that control the degree of enhancement (ranging from 0 to 1).
Specifically, we set $\alpha_{sp}=0.5$ and $\alpha_{ns}=0.9$, which are determined empirically through ablation studies. Detailed analyses are provided in Appendix~\ref{hyper}.

The MAFB performs modality-aware signal fusion between separated signals and raw audio. This design balances modality enhancement and signal fidelity. When separated signals contain artifacts, mixing in original audio preserves natural acoustics and improves LALMs' robustness.

\subsection{Audio Separator}
In FTL, we employ an audio separation model to separate speech from non-speech components. To achieve this, we first consider existing state-of-the-art (SOTA) pre-trained models: SE-Mamba (\textbf{SEM})~\cite{se-mamba} and SAM-Audio (\textbf{SAM})~\cite{sam}. 

Specifically, SEM is a GAN~\cite{gan}-based speech enhancement model; the enhanced speech is first estimated from the mixture, and the non-speech signal is obtained by subtracting the enhanced speech from the mixture. SAM is a generative separation model that simultaneously estimates both the target and residual stems from an audio mixture, conditioned on text or visual prompts; we use a text prompt with the content ``\textit{speech}''.

However, SEM is trained with speech enhancement objectives instead of separation, and SAM may generate signal components not present in the raw audio, which can potentially mislead downstream audio understanding tasks.
Therefore, we develop \textbf{SNSep}, a separator specialized for speech and non-speech separation, which operates in the short-time Fourier transform domain using a masking-based approach. 
Specifically, we adopt the separation network from AudioSep~\cite{audiosep,yin-audiosep} as the backbone. 
SNSep has a dual-decoder architecture: one decoder reconstructs the speech track, while a parallel decoder independently extracts the non-speech track.

\section{Experimental Setups}

\subsection{Detailed Configurations}
\noindent \textbf{SNSep Training:} For training SNSep, we sample 50 hours of speech from LJSpeech, Librispeech, VoxPopuli, and GigaSpeech training sets~\cite{ljspeech,librispeech,voxpopuli,gigaspeech} and 50 hours of non-speech from VocalSound, VGGSound,
CochlScene, AudioSet, FSD50K, and UrbanSound8K training sets~\cite{vocalsound,vggsound,cochlscene,audioset,fsd50k,urbansound8k}. 
During training, a speech and a non-speech sample are mixed with an SNR randomly selected from -10 \,dB to 10 \,dB as the input, and all audio samples are resampled to 16 kHz.
For other configurations, we follow the previous work~\cite{audiosep}.

\noindent \textbf{Modality Router and LALM:} For the modality router, we use Qwen3-8B~\cite{qwen3} and ChatGPT5.2. For the LALM, we adopt Audio Flamingo 3 (\textbf{AF3})~\cite{af3}, Fun-Audio-Chat (\textbf{FAC})~\cite{funaudiochat}, and Qwen3-Omni (\textbf{Q3O})~\cite{qwen3-omni}.

\subsection{Evaluation}
We investigate the effectiveness of FTL on \textbf{audio perception} and \textbf{audio reasoning} tasks.
Audio perception tasks primarily assess the model's ability to recognize and understand low-level acoustic events and spoken content, while audio reasoning tasks require higher-level semantic inference and compositional reasoning over auditory objects. Through these experiments, we aim to investigate whether FTL consistently improves both perceptual understanding and reasoning capabilities in LALMs.

\noindent \textbf{Audio Perception:} We use SSEU-Bench to evaluate audio perception performance in noisy conditions, where each audio sample is mixed with a speech and a non-speech sound with a specific SNR.
Two classic tasks in speech and environmental sound domains are included : Automatic Speech Recognition (ASR) and Audio Tagging (AT). 
For ASR, the LALM is required to output the spoken content of the speaker. For AT, we require the LALM to detect non-speech sound events within the audio. Detailed instructions used for the two tasks are provided in Appendix~\ref{prompt:user}.

\begin{table}[t]
\centering
\caption{ASR performance of LALMs on SSEU-Bench. Metric is WER(\%). ``SNR-Speech'' refers to speech to non-speech ratio, where ``SNR-Speech=$+\infty$'' means pure speech.}
\renewcommand\arraystretch{0.5}{
\setlength{\tabcolsep}{1.3mm}{
\scalebox{0.95}{
{\footnotesize
\begin{tabular}{c|c|cccccc}
\toprule
\multirow{2}{*}{LALM} & \multirow{2}{*}{FTL} & \multicolumn{6}{c}{SNR-Speech (\,dB)} \\
& & $+\infty$ & 10 & 5 & 0 & -5 & -10\\
\midrule
\multirow{2}{*}{AF3} & \ding{55} & 2.18 & 2.71 & 3.27 & 4.73 & 10.45 & 27.45 \\
& \ding{52} & \textbf{2.17} & \textbf{2.66} & \textbf{3.20} & \textbf{4.61} & \textbf{9.83} & \textbf{25.39}\\
\midrule
\multirow{2}{*}{FAC} & \ding{55} & \textbf{2.61} & 3.38 & 3.99 & 5.75 & 12.54 & 31.67\\
& \ding{52} & \textbf{2.61} & \textbf{3.24} & \textbf{3.86} & \textbf{5.44} & \textbf{11.54} & \textbf{28.41}\\
\midrule
\multirow{2}{*}{Q3O} & \ding{55} & \textbf{2.16} & \textbf{2.31} & 2.56 & 3.64 & 7.04 & 20.42 \\
& \ding{52} & 2.23 & \textbf{2.31} & \textbf{2.49} & \textbf{3.38} & \textbf{5.97} & \textbf{18.61} \\
\bottomrule
\end{tabular}}
}
}
}
\label{tab:exp-1}
\end{table}

\begin{table}[t]
\centering
\caption{AT performance of LALMs on SSEU-Bench. Metric is mAP(\%). ``SNR-Non-Speech'' refers to non-speech to speech ratio, where ``SNR-Non-Speech=$+\infty$'' means pure non-speech.}
\renewcommand\arraystretch{0.5}{
\setlength{\tabcolsep}{1.3mm}{
\scalebox{0.95}{
{\footnotesize
\begin{tabular}{c|c|cccccc}
\toprule
\multirow{2}{*}{LALM} & \multirow{2}{*}{FTL} & \multicolumn{6}{c}{SNR-Non-Speech (\,dB)} \\
& & $+\infty$ & 10 & 5 & 0 & -5 & -10\\
\midrule
\multirow{2}{*}{AF3} & \ding{55} & 38.80 & 36.18 & 34.56 & 31.00 & 28.86 & 27.36\\
& \ding{52} & \textbf{39.28} & \textbf{39.26} & \textbf{38.95} & \textbf{38.19} & \textbf{34.94} & \textbf{31.56}\\
\midrule
\multirow{2}{*}{FAC} & \ding{55} & 36.34 & 21.27 & 18.33 & 17.54 & 16.98 & 16.34\\
& \ding{52} & \textbf{36.64} & \textbf{31.97} & \textbf{30.32} & \textbf{27.88} & \textbf{24.89} & \textbf{20.75}\\
\midrule
\multirow{2}{*}{Q3O} & \ding{55} & \textbf{44.43} & 39.75 & 38.12 & 34.84 & 33.20 & 31.33\\
& \ding{52} & 44.27 & \textbf{43.48} & \textbf{42.25} & \textbf{40.32} & \textbf{39.20} & \textbf{37.27}\\
\bottomrule
\end{tabular}}
}
}
}
\label{tab:exp-2}
\vspace{-1em}
\end{table}

\noindent \textbf{Audio Reasoning:} MMAU-Pro~\cite{mmau-pro} is a widely used audio reasoning benchmark, which comprises various audio-based question–answer (QA) pairs. 
However, MMAU-Pro does not provide specific SNRs for speech and non-speech components within an audio sample.
Therefore, we curate a new subset of MMAU-Pro with controllable SNR conditions, i.e., \textbf{MMAU-Pro-Ctrl}.

Specifically, we manually collect 130 speech- and 130 non-speech-QAs from MMAU-Pro. For the speech-QA subset, we ensure that the audio consists of clean speech (4s to 300s) with questions explicitly target speech content. Conversely, the non-speech subset contains non-speech audio (5s to 293s) with questions. To simulate realistic noisy speech-QAs, we utilize the non-speech samples as the noise. For each pair, noise shorter than the speech is randomly inserted, whereas longer noise is cropped to match its duration. The same mixing protocol is used for non-speech QAs, with speech treated as noise. Following SSEU-Bench, we range the SNR from 10\,dB to -10\,dB.

\textbf{Metrics:} We use \textbf{Word Error Rate (WER)} to evaluate ASR. 
For AT, we use \textbf{mean Average Precision (mAP)} for evaluation.
For reasoning tasks, we follow MMAU-Pro and use the averaged accuracy for evaluation, denoted as \textbf{QA-ACC}. In addition, we report the \textbf{Correct Rate (CR)} to measure the performance of the modality router, which is defined as the proportion of samples where the target audio modality is correctly predicted. 

\section{Results and Discussions} 

\begin{table}[t]
\centering
\caption{Reasoning performance (QA-ACC(\%)) on MMAU-Pro-Ctrl with different modality routers (LLM: Qwen3-Omni).} 
\renewcommand\arraystretch{0.5}{
\setlength{\tabcolsep}{0.9mm}{
\scalebox{0.95}{
{\footnotesize
\begin{tabular}{c|c|c|cccccc}
\toprule
\multicolumn{9}{c}{\textbf{Speech Reasoning}} \\
\midrule
\multirow{2}{*}{FTL} & \multirow{2}{*}{Modality Router} & \multirow{2}{*}{CR(\%)} & \multicolumn{6}{c}{SNR-Speech (\,dB)} \\
& & &$+\infty$& 10 & 5 & 0 & -5 & -10 \\
\midrule
\ding{55} & - &  - & \textbf{75.4} & 75.4 & \textbf{75.4} & 74.6 & 73.1 & 70.0 \\
\ding{52} & Qwen3-8B & 23.8 & \textbf{75.4} & 74.6 & 74.6 & 73.8 & \textbf{74.6} & 70.0\\
\ding{52} & ChatGPT5.2 & 88.5 & \textbf{75.4} & \textbf{76.2} & \textbf{75.4} & \textbf{75.4} & \textbf{74.6} & \textbf{73.1}\\
{\color{gray!80} \ding{52}} & {\color{gray!80} GroundTruth} & {\color{gray!80}100.0} & {\color{gray!80}76.2} & {\color{gray!80}75.4} & {\color{gray!80}75.4} & {\color{gray!80}75.4} & {\color{gray!80}73.8} & {\color{gray!80}72.3}\\
\midrule
\multicolumn{9}{c}{\textbf{Non-Speech Reasoning}} \\
\midrule
\multirow{2}{*}{FTL} & \multirow{2}{*}{Modality Router} & \multirow{2}{*}{CR(\%)} & \multicolumn{6}{c}{SNR-Non-Speech (\,dB)} \\
& & & $+\infty$& 10 & 5 & 0 & -5 & -10 \\
\midrule
\ding{55} & - & -& \textbf{43.1} & 35.4 & 38.5 & 37.7 & 36.2 & 34.6 \\
\ding{52} & Qwen3-8B & 0.0 & \textbf{43.1} & 35.4 & 38.5 & 37.7 & 36.2 & 34.6 \\
\ding{52} & ChatGPT5.2 & 47.7 & 41.5 & \textbf{42.3} & \textbf{43.1} & \textbf{40.0} & \textbf{39.2} & \textbf{38.5} \\
{\color{gray!80} \ding{52}} & {\color{gray!80} GroundTruth} & {\color{gray!80}100.0} & {\color{gray!80}42.3} & {\color{gray!80}42.3} & {\color{gray!80}40.8} & {\color{gray!80}40.0} & {\color{gray!80}39.2} & {\color{gray!80}38.5}\\
\bottomrule

\end{tabular}}
}
}
}
\label{tab:exp-3}
\vspace{-1em}
\end{table}

\noindent \textbf{Effectiveness of FTL on Audio Perception:} Table~\ref{tab:exp-1} and Table~\ref{tab:exp-2} present the performance of LALMs on ASR and AT tasks, where Qwen3-8B is applied as the modality router. 

For ASR, as shown in Table 1, FTL effectively reduces the WER of all evaluated LALMs, especially under low-SNR conditions where non-speech interference becomes dominant. For example, under the \textit{SNR-Speech} of $-10$ dB, FTL reduces the WER of AF3 from 27.45\% to 25.39\%, and also improves FAC and Q3O by a similar margin.

For AT, Table 2 shows that FTL consistently improves mAP across different \textit{SNR-Non-Speech} settings. The improvements are particularly notable when the target non-speech signals are contaminated by strong speech interference. For instance, under the \textit{SNR-Non-Speech} of $-10$ dB, FTL improves the mAP of FAC from 16.34\% to 20.75\%.

These results suggest that our proposed FTL provides a general and effective mechanism for task-adaptive audio enhancement, benefiting both speech-centric and non-speech-centric audio perception tasks.

\noindent \textbf{Effectiveness of FTL on Audio Reasoning:} Table~\ref{tab:exp-3} presents the reasoning performance of FTL on MMAU-Pro-Ctrl under different modality routers and SNR conditions.

For speech reasoning, FTL generally improves QA-ACC under noisy conditions when accurate modality routing is available.
For non-speech reasoning, FTL with ChatGPT5.2 and ground-truth routing also improves performance across most SNR settings.
It should be noted that, Qwen3-8B achieves a CR of 0\% because it consistently predicts ``mixture'', resulting in the original mixed audio being fed into the LALMs and thus providing no performance gain.

A surprising observation is that a better router does not always bring better reasoning performance.
In some cases, the ground-truth router performs comparably to or even slightly worse than ChatGPT5.2 despite perfect routing accuracy.
This suggests that, compared with low-level perception tasks, audio reasoning is more sensitive to signal distortion and contextual completeness, making the effectiveness of modality-aware enhancement less deterministic.

\noindent \textbf{Impact of Audio Separator:} In previous experiments, we use SNSep as the default separator. In Table 4, we further investigate the impact of different audio separators within the proposed FTL framework. In addition, the SDR performance of different separators is provided in Appendix~\ref{ab:sep}, where SNSep and SEM achieve comparable separation quality, significantly outperforming SAM.

As shown in Table~\ref{tab:exp-4}, both SEM and SNSep consistently improve ASR and AT performance compared with the vanilla model, whereas SAM often leads to degraded ASR results.
An interesting observation is that although SNSep achieves slightly better separation performance than SEM in terms of SDR, it does not yield better ASR improvements.
This finding suggests that cleaner separation does not necessarily lead to better speech understanding for LALMs. Excessively aggressive separation may remove subtle acoustic cues or introduce distortions that negatively affect downstream perception. We provide a real-case analysis in Appendix~\ref{ab:real-case} to further illustrate this phenomenon.

\noindent \textbf{Performance on Real Mixtures:} Since real mixtures lack ground-truth SNRs, we provide qualitative demonstrations of FTL on real-world mixtures on our project page\textsuperscript{\ref{fn:page}}.
Results show that FTL can also effectively improve audio understanding in real-world mixtures.

\section{Conclusions}

In this work, we propose FTL, an easy but efficient audio enhancement framework for noise-robust LALMs.
Results show that FTL improves both audio perception and reasoning performance, especially under high-noise conditions, providing practical guidelines for deploying LALMs in real-world noisy scenarios.
Despite its effectiveness, FTL applies a frozen LLM for modality routing; future work will study adaptive fusion and routing to improve robustness.

\begin{table}[t]
\centering
\caption{Performance of Audio Flamingo 3 with different audio separators in FTL on SSEU-Bench.}
\renewcommand\arraystretch{0.5}{
\setlength{\tabcolsep}{1.6mm}{
\scalebox{0.95}{
{\footnotesize
\begin{tabular}{c|c|ccccc}
\toprule
\multicolumn{7}{c}{\textbf{ASR Performance, Metric: WER (\%)}} \\
\midrule
\multirow{2}{*}{FTL} & \multirow{2}{*}{Sep} & \multicolumn{5}{c}{SNR-Speech (\,dB)} \\
& & 10 & 5 & 0 & -5 & -10 \\
\midrule
\ding{55} & - & 2.71 & 3.27 & 4.73 & 10.45 & 27.45 \\
\ding{52} & SAM & 2.83 & 3.31 & 4.93 & 10.40 & 28.72 \\
\ding{52} & SEM & \textbf{2.62} & \textbf{3.03} & \textbf{4.08} & \textbf{8.07} & \textbf{23.83} \\
\ding{52} & SNSep & 2.66 & 3.20 & 4.61 & 9.83 & 25.39 \\
\midrule
\multicolumn{7}{c}{\textbf{AT Performance, Metric: mAP (\%)}} \\
\midrule
\multirow{2}{*}{FTL} & \multirow{2}{*}{Sep} & \multicolumn{5}{c}{SNR-Non-Speech (\,dB)} \\
& & 10 & 5 & 0 & -5 & -10 \\
\midrule
\ding{55} & - & 36.18 & 34.56 & 31.00 & 28.86 & 27.36 \\
\ding{52} & SAM & 36.61 & 37.89 & 35.56 & 33.19 & 31.98\\
\ding{52} & SEM & 38.36 & 38.52 & 36.74 & \textbf{35.12} & \textbf{33.67}\\
\ding{52} & SNSep & \textbf{39.26} & \textbf{38.95} & \textbf{38.19} & 34.94 & 31.56\\
\bottomrule
\end{tabular}}
}
}
}
\label{tab:exp-4}
\vspace{-1em}
\end{table}


\section{Acknowledgments}
This work was supported by the National Research Foundation of Korea (NRF) grant (No. RS-2024-00337945); and the BK21 FOUR program through the NRF grant funded by the Ministry of Education of Korea government (MOE). This work was supported by Artificial intelligence industrial convergence cluster development project funded by the Ministry of Science and ICT(MSIT, Korea)\&Gwangju Metropolitan City.

\nocite{langley00}

\bibliography{example_paper}
\bibliographystyle{icml2026}

\newpage
\appendix
\onecolumn

\setcounter{table}{0}
\renewcommand{\thetable}{A\arabic{table}}

\setcounter{figure}{0}
\renewcommand{\thefigure}{A\arabic{figure}}

\section{Prompts}
\subsection{Prompt for the LLM-based modality router in FTL}
\label{prompt:agent}

You are an expert in audio understanding and multimodal reasoning. Your task is to decide what audio input should be provided to a Large Audio Language Model (LALM) in order to best accomplish a user’s instruction. 

The audio has been separated into two tracks: speech: contains spoken voice content only; non-speech: contains non-speech acoustic events only. Mixture refers to the original unseparated audio.

You should select the input that maximizes task-relevant information, based on the user’s instruction. 

Guidelines: 

1. You should ONLY choose `speech' when speech information alone is clearly sufficient to solve the task, AND non-speech provides no meaningful additional information. 

2. You should ONLY choose `non-speech' when non-speech audio alone is clearly sufficient to solve the task, AND speech provides no meaningful additional information. 

3. In ALL other cases, including uncertainty, partial usefulness of both modalities, or when you cannot strictly rule out one modality, you MUST choose `mixture'. 

Additional Domain Rules: - Speech is required for linguistic content, speaker intent, emotion, or dialogue understanding. - Non-speech includes environmental sounds and vocal non-linguistic sounds (e.g., laughter, sneeze, cough). 

Respond with only one word: speech, non-speech, or mixture. Do not provide explanations. 

User Instruction: [\textbf{the user's instruction}].

\subsection{User's Instructions}
\label{prompt:user}
\subsubsection{Automatic Speech Recognition (ASR) Task}
Transcribe the speech into text, without any further explanation.
\subsubsection{Audio Tagging (AT) Task}
You are an expert in sound events classification. I will give you an audio recording. Please carefully analyze the sound events in this audio. Ignore speech and focus only on non-speech sound events. Output only one line, no explanations. List events detected in the audio, separated by a semicolon and a space. If no event is detected, output: None.

\section{Ablations of Hyper-parameters}
\label{hyper}

\begin{figure}[htbp]
\centering
\includegraphics[width=0.7\columnwidth]{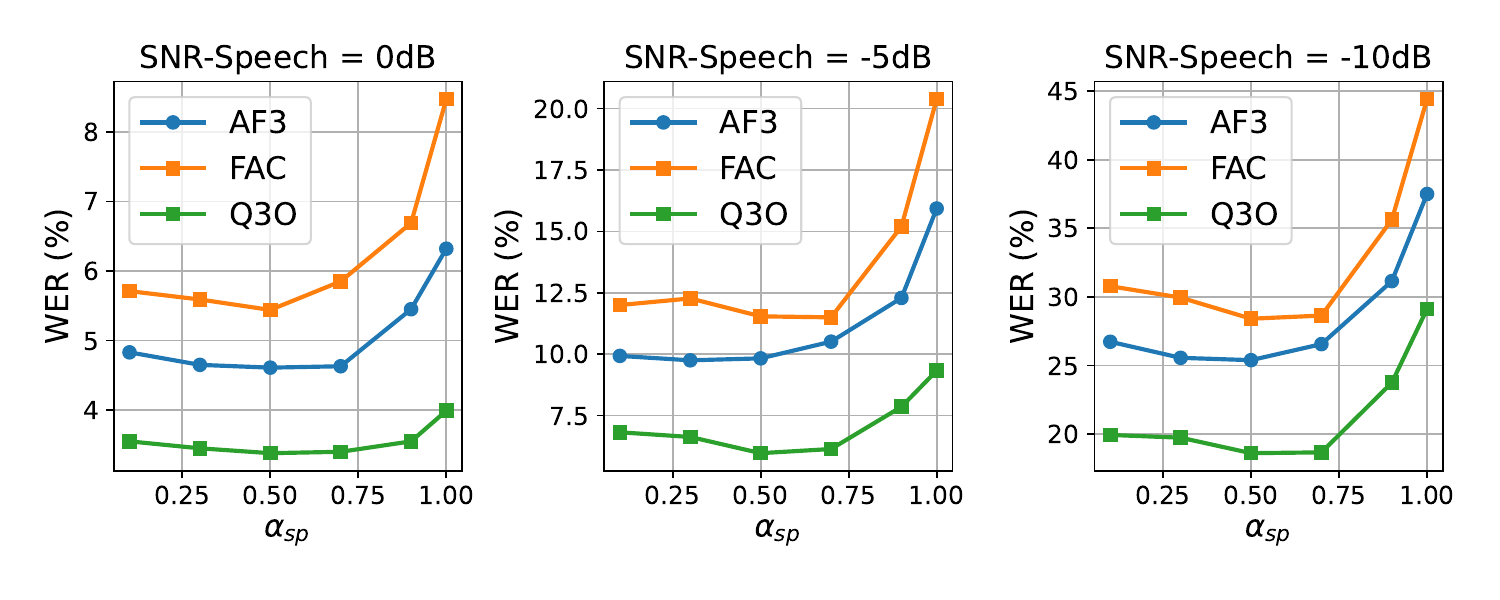}
\vspace{-1.5em}
\caption{Impact of $\alpha_{sp}$ on ASR task of SSEU-Bench.}
\label{fig:alpha_sp}
\vspace{-1.5em}
\end{figure}

In this section, we analyze the impact of the fusion coefficients $\alpha_{sp}$ and $\alpha_{ns}$ on the performance of FTL, and explain how the default values ($\alpha_{sp}=0.5$ and $\alpha_{ns}=0.9$) are determined. Table~\ref{tab:ablation_alpha} reports the ASR and AT results of three LALMs (AF3, FAC, and Q3O) under different $\alpha_{sp}$ and $\alpha_{ns}$ values on SSEU-Bench. Figure~\ref{fig:alpha_sp} further visualizes the impact of $\alpha_{sp}$ on ASR at three representative SNR-Speech conditions.

\textbf{Impact of $\alpha_{sp}$ on ASR:} As shown in Table~\ref{tab:ablation_alpha} and Figure~\ref{fig:alpha_sp}, the WER of all three LALMs exhibits a clear U-shaped trend with respect to $\alpha_{sp}$, especially under low-SNR conditions. When $\alpha_{sp}=1.0$, i.e., the LALM directly consumes the separated speech signal, the WER becomes substantially worse than the vanilla baseline. For example, at SNR-Speech~$=-10$~dB, $\alpha_{sp}=1.0$ degrades the WER of AF3 from $27.45\%$ to $37.50\%$, and FAC from $31.67\%$ to $44.41\%$. This degradation indicates that, although fully replacing the input with the separated speech effectively removes non-speech interference, the separator inevitably introduces artifacts and spectral distortions that mislead the acoustic perception of LALMs. On the other hand, when $\alpha_{sp}=0.1$, the enhanced signal is dominated by the raw mixture, providing only marginal suppression of non-speech interference and thus limited improvement over the baseline. The setting of $\alpha_{sp}=0.5$ achieves the best overall balance: it removes a substantial portion of non-speech interference while retaining sufficient natural acoustic context from the raw audio to preserve signal fidelity. As a result, $\alpha_{sp}=0.5$ consistently yields the lowest WER across all three LALMs and most SNR levels (e.g., AF3 from $27.45\%$ to $25.39\%$ and Q3O from $20.42\%$ to $18.61\%$ at SNR-Speech~$=-10$~dB), and is therefore selected as the default value.

\begin{table*}[t]
\centering
\caption{Performance of different LALMs on SSEU-Bench. ``SNR-Speech'' denotes the speech-to-non-speech ratio; ``SNR-Non-Speech'' refers to non-speech to speech ratio.}
\renewcommand\arraystretch{1.3}{
\setlength{\tabcolsep}{2mm}{
\scalebox{0.95}{
{\footnotesize
\begin{tabular}{c|c|c|cccccc|c|cccccc}
\toprule
\multirow{3}{*}{LALM} & \multirow{3}{*}{FTL} & \multirow{3}{*}{$\alpha_{sp}$} & \multicolumn{6}{c|}{\textbf{ASR Performance, Metric: WER (\%)}} & \multirow{3}{*}{$\alpha_{ns}$} & \multicolumn{6}{c}{\textbf{AT Performance, Metric: mAP (\%)}}\\
\cline{4-9}
\cline{11-16}
& & & \multicolumn{6}{c|}{SNR-Speech (\,dB)} & & \multicolumn{6}{c}{SNR-Non-Speech (\,dB)}\\
& & & $+\infty$ & 10 & 5 & 0 & -5 & -10 & & $+\infty$ & 10 & 5 & 0 & -5 & -10\\
\midrule
\multirow{5}{*}{AF3} & \ding{55} & - & 2.18 & 2.71 & 3.27 & 4.73 & 10.45 & 27.45 & -  & 38.80 & 36.18 & 34.56 & 31.00 & 28.86 & 27.36\\
& \ding{52} & 1.0 & 2.21 & 3.13 & 3.93 & 6.32 & 15.93 & 37.50 & 1.0 & 39.22 & 38.55 & 37.43 & 36.44 & 34.86 & \textbf{31.94}\\
& \ding{52} & 0.9 & \textbf{2.15}  & 2.89 & 3.49 & 5.45 & 12.29 & 31.15 & 0.9& \textbf{39.28} & \textbf{39.26} & \textbf{38.95} & \textbf{38.19} & \textbf{34.94} & 31.56\\
& \ding{52} & 0.5 & 2.17 & \textbf{2.66} & \textbf{3.20} & \textbf{4.61} & \textbf{9.83} & \textbf{25.39} & 0.5 & 39.16 & 36.52 & 35.34 & 32.92 & 31.24 & 29.29\\
& \ding{52} & 0.1 & 2.16 & 2.70  & 3.43  & 4.63  & 9.93  & 26.73 & 0.1 & 39.19 & 36.06 & 33.01 & 31.51 & 29.34 & 27.70\\
\midrule
\multirow{5}{*}{FAC} & \ding{55} & - & 2.61& 3.38 & 3.99 & 5.75 & 12.54 & 31.67 & - & 36.34 & 21.27 & 18.33 & 17.54 & 16.98 & 16.34\\
& \ding{52} & 1.0 & 2.82 & 3.66 & 4.94 & 8.47 & 20.38 & 44.41 & 1.0 & 36.34 & \textbf{33.22} & \textbf{31.73} & \textbf{32.32} & \textbf{31.77} & \textbf{29.30}\\
& \ding{52} & 0.9 & \textbf{2.58}& 3.37 & 4.26 & 6.69 & 15.20 & 35.63 & 0.9 & \textbf{36.64} & 31.97 & 30.32 & 27.88 & 24.89 & 20.75\\
& \ding{52} & 0.5 & 2.61 & \textbf{3.24} & \textbf{3.86} & \textbf{5.44} & \textbf{11.54} & \textbf{28.41} & 0.5 & 36.16 & 26.62 & 21.10 & 18.43 & 17.74 & 17.39\\
& \ding{52} & 0.1 & \textbf{2.58} & 3.32 & 3.91 & 5.71 & 12.00 & 30.78 & 0.1 & 36.60 & 21.80 & 18.58 & 17.78 & 17.42 & 16.31\\
\midrule
\multirow{5}{*}{Q3O} & \ding{55} & - & 2.16 & 2.31 & 2.56 & 3.64 & 7.04 & 20.42 & - & 44.43 & 39.75 & 38.12 & 34.84 & 33.20 & 31.33\\
& \ding{52} & 1.0 & \textbf{2.05} & \textbf{2.18} & 2.66 & 3.99 & 9.33 & 29.12 & 1.0 & \textbf{44.66} & \textbf{43.87} & \textbf{43.46} & \textbf{42.01} & \textbf{39.94} & \textbf{37.30}\\
& \ding{52} & 0.9& 2.14 & 2.45 & 2.66 & 3.55 & 7.86 & 23.75 & 0.9 & 44.27 & 43.48 & 42.25 & 40.32 & 39.20 & 37.27\\
& \ding{52} & 0.5& 2.23 & 2.31 & \textbf{2.49} & \textbf{3.38} & \textbf{5.97} & \textbf{18.61}& 0.5 & 44.38 & 40.49 & 38.65 & 37.58 & 34.76 & 32.97\\
& \ding{52} & 0.1& 2.20 & 2.38 & 2.58 & 3.55 & 6.82 & 19.94& 0.1 & 44.55 & 40.31 & 37.83 & 35.88 & 33.13 & 30.98\\
\bottomrule

\end{tabular}}
}
}
}
\label{tab:ablation_alpha}
\vspace{-1em}
\end{table*}

\textbf{Impact of $\alpha_{ns}$ on AT:} Unlike ASR, the AT performance generally improves as $\alpha_{ns}$ increases. As shown in Table~\ref{tab:ablation_alpha}, when $\alpha_{ns}=1.0$, the LALM receives the purely separated non-speech signal and achieves strong AT performance across all LALMs and SNR conditions. For instance, at SNR-Non-Speech~$=-10$~dB, $\alpha_{ns}=1.0$ improves the mAP of AF3 from $27.36\%$ to $31.94\%$, and Q3O from $31.33\%$ to $37.30\%$. Setting $\alpha_{ns}=0.9$ produces results highly comparable to $\alpha_{ns}=1.0$, and even surpasses it in several cases (e.g., AF3 achieves $39.26\%$ mAP at SNR-Non-Speech~$=10$~dB with $\alpha_{ns}=0.9$, higher than $38.55\%$ with $\alpha_{ns}=1.0$). In contrast, smaller $\alpha_{ns}$ values ($0.5$ and $0.1$) yield notably weaker AT performance, since the enhanced signal still contains a large portion of speech interference that distracts the LALMs from non-speech.

Although $\alpha_{ns}=1.0$ slightly outperforms $\alpha_{ns}=0.9$ on the AT task in some settings, we adopt $\alpha_{ns}=0.9$ as the default value for the following robustness consideration. When $\alpha_{ns}=1.0$, the enhanced signal consists exclusively of the separated non-speech component, meaning that the speech track is completely discarded. In practice, the modality router is not always perfect: if the router misjudges a speech-related instruction as ``non-speech'', such an aggressive setting would eliminate all spoken content from the input, leading to a catastrophic failure on the downstream task. Retaining $10\%$ of the raw audio ($\alpha_{ns}=0.9$) ensures that even under occasional routing errors, the speech information is partially preserved, preventing complete loss of task-relevant content. This conservative design trades a small amount of AT performance for substantially better robustness against router errors, which we believe is a favorable trade-off for real-world deployment.

\textbf{Summary:} Based on the above analysis, we adopt $\alpha_{sp}=0.5$ and $\alpha_{ns}=0.9$ as the default hyper-parameters in FTL. These values are chosen not only for their strong empirical performance, but also to jointly balance enhancement strength, signal fidelity, and robustness against potential router misjudgments.

\section{Performance of Audio Separators}
\label{ab:sep}

\begin{figure}[htbp]
\centering
\includegraphics[width=0.7\columnwidth]{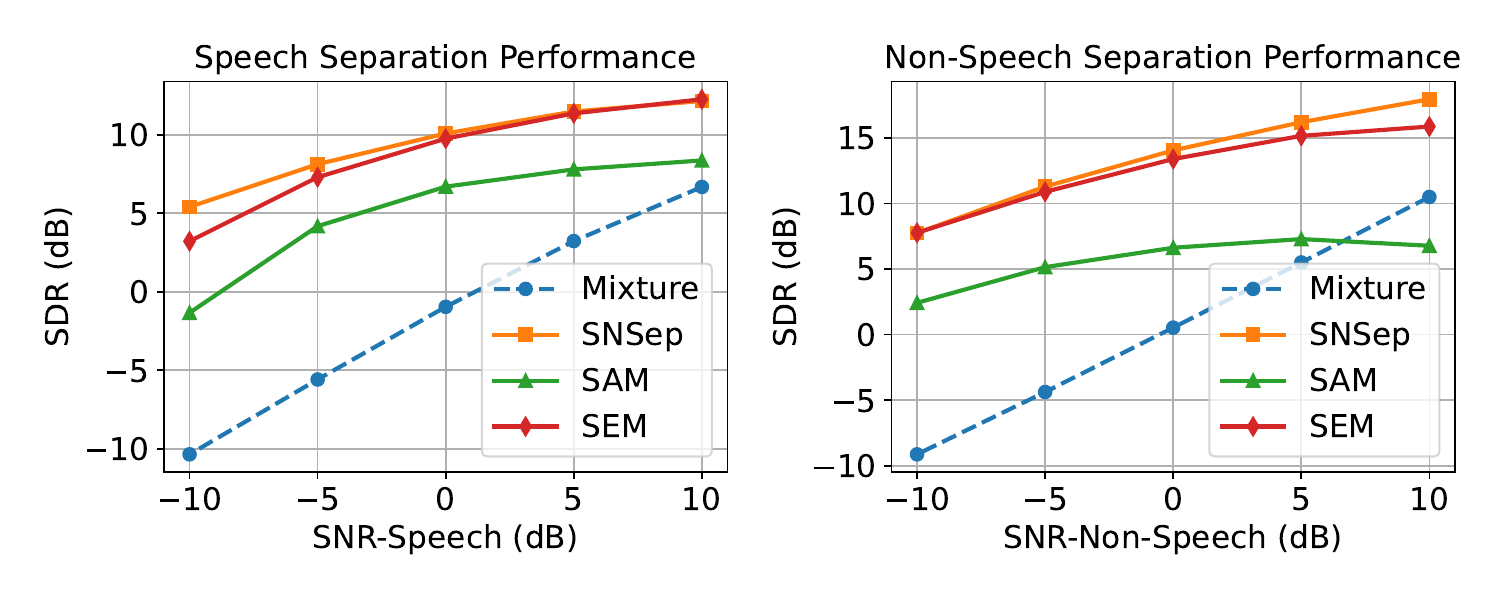}
\caption{Speech and non-speech separation performance of different separators on SSEU-Bench.}
\label{fig:separator_sdr}
\end{figure}

In this section, we evaluate the separation quality of three audio separators considered in FTL, i.e., SNSep, SAM, and SEM. Figure~\ref{fig:separator_sdr} reports the Signal-to-Distortion Ratio (SDR) of each separator on the speech and non-speech tracks of SSEU-Bench, under varying SNR conditions. As a reference, we also include the SDR of the raw mixture, which reflects the difficulty of the input.

\textbf{Overall Comparison:} As shown in Figure~\ref{fig:separator_sdr}, all three separators substantially outperform the raw mixture baseline on both speech and non-speech tracks, confirming that explicit separation provides meaningful signal-level improvement for both target modalities. Among the three separators, SNSep and SEM consistently achieve the highest SDR across all SNR levels, while SAM lags behind by a noticeable margin. For instance, at SNR-Speech~$=-10$~dB, SNSep and SEM yield speech SDRs above $3$~dB, whereas SAM only reaches around $-1$~dB. A similar trend is observed on the non-speech track, where SNSep and SEM clearly outperform SAM by roughly $5$~dB at low-SNR conditions.

\textbf{Comparison between SNSep and SEM:} SNSep and SEM achieve highly comparable separation quality in terms of SDR, with SNSep slightly better in most SNR conditions, particularly on the non-speech track at low SNRs. This is consistent with our design motivation: SEM is a speech enhancement model that estimates non-speech by subtracting the enhanced speech from the mixture, which inevitably propagates speech-estimation errors into the non-speech branch. In contrast, SNSep adopts a dedicated dual-decoder architecture that independently reconstructs the speech and non-speech tracks, leading to a more balanced separation across both modalities.

\textbf{Discussion:} The SDR results above explain the trends observed in Table~\ref{tab:exp-4} of the main paper. Specifically, both SNSep and SEM achieve consistent improvements on ASR and AT when used in FTL, while SAM occasionally degrades the ASR performance due to its inferior separation quality. Notably, although SNSep yields slightly higher SDR than SEM, it does not always translate into better downstream ASR performance. This observation indicates that signal-level metrics such as SDR are not perfectly aligned with the perceptual preferences of LALMs, and that excessive separation may remove subtle acoustic cues useful for downstream understanding. We adopt SNSep as the default separator in FTL because of its overall balanced separation quality across both speech and non-speech tracks, while leaving the joint optimization of separation and downstream LALM perception as an interesting direction for future work.

\section{Real-Case Analysis}
\label{ab:real-case}

\begin{figure}[htbp]
\centering
\includegraphics[width=0.7\columnwidth]{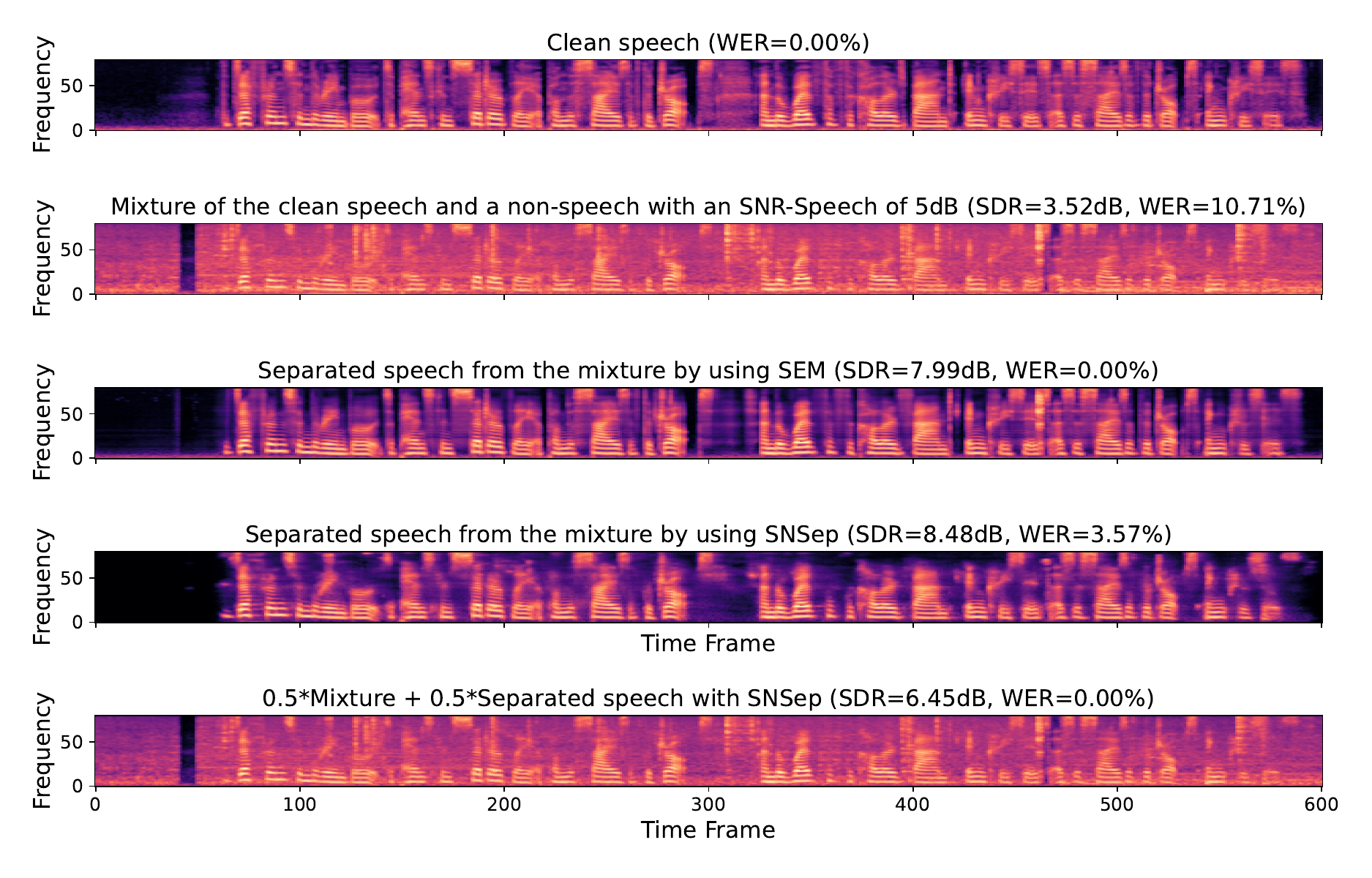}
\caption{ASR demonstration (mel spectrogram): Each sample is fed into Audio Flamingo 3 to perform ASR.}
\label{fig:asr_demo}
\end{figure}

To better understand why a separator with higher SDR does not always lead to better downstream ASR performance, we present a representative case study in Figure~\ref{fig:asr_demo}. The example consists of a clean speech utterance and a non-speech sound mixed at SNR-Speech~$=5$~dB. We feed five different versions of this sample into Audio Flamingo 3 (AF3) and report both the SDR (against the clean speech reference) and the corresponding WER.

\textbf{SDR is not Always Aligned with ASR Performance:} As shown in Figure~\ref{fig:asr_demo}, the raw mixture yields a low SDR of $3.52$~dB and causes AF3 to produce a WER of $10.71\%$. After separation, both SEM and SNSep substantially increase the SDR ($7.99$~dB and $8.48$~dB, respectively), confirming that they effectively remove the non-speech component at the signal level. However, the downstream ASR results tell a different story: SEM enables AF3 to perfectly transcribe the utterance ($\text{WER}=0.00\%$), whereas SNSep, despite having the highest SDR, still leads to a non-trivial $\text{WER}=3.57\%$. This counter-intuitive result aligns with our observation in Section~4 of the main paper, namely that signal-level separation quality does not directly translate into perceptual benefits for LALMs.

\textbf{Why Does SNSep Hurt ASR?} A closer inspection of the mel spectrograms reveals the underlying cause. Compared with the clean speech reference and the SEM-separated output, the SNSep-separated speech exhibits a clearly attenuated low-frequency band at the bottom of the mel spectrogram (highlighted by the missing horizontal energy strip in Figure~\ref{fig:asr_demo}). This region typically carries vocal fundamental frequencies and low-order harmonics that are crucial for natural speech perception. Although removing such energy bands helps suppress residual non-speech interference (and thus yields a slightly higher SDR), it also makes the resulting waveform acoustically unnatural and out-of-distribution from the perspective of the pre-trained LALM. As a result, AF3 misrecognizes some words despite the higher SDR. This phenomenon illustrates that overly aggressive separation can sacrifice acoustic naturalness, which is more important than raw signal fidelity for downstream LALM perception.

\textbf{How MAFB Mitigates the Issue:} The bottom panel of Figure~\ref{fig:asr_demo} shows the output of our MAFB with $\alpha_{sp}=0.5$, i.e., the linear combination of the raw mixture and the SNSep-separated speech. Although this fusion lowers the SDR to $6.45$~dB (since the mixture re-introduces a small amount of non-speech interference), it restores the missing low-frequency band and recovers the natural acoustic structure of speech. Crucially, AF3 now achieves a perfect $\text{WER}=0.00\%$ on this fused signal. This case study provides direct empirical support for the design of MAFB: by mixing the separated signal with the raw audio, MAFB compensates for the artifacts and missing acoustic cues introduced by aggressive separation, achieving a better trade-off between noise suppression and signal naturalness than relying on the separator output alone.


\end{document}